\documentclass[%
 reprint,
 superscriptaddress,
 amsmath,amssymb,
 aps,
]{revtex4-2}

\usepackage{graphicx}
\usepackage{dcolumn}
\usepackage{bm}
\usepackage{color} 

\begin{document}


\title{Finite-size effects on the convergence time in continuous-opinion dynamics}

\author{Hang-Hyun Jo}
\email{h2jo@catholic.ac.kr}
\affiliation{%
 Department of Physics, The Catholic University of Korea, Bucheon 14662, Republic of Korea
}%

\author{Naoki Masuda}
\affiliation{
Department of Mathematics, State University of New York at Buffalo 14260-2900, USA
}
\affiliation{
Computational and Data-Enabled Science and Engineering Program, State University of New York at Buffalo 14260-5030, USA
}

\date{\today}

\begin{abstract}
We study finite-size effects on the convergence time in a continuous-opinion dynamics model. In the model, each individual's opinion is represented by a real number on a finite interval, e.g., $[0,1]$, and a uniformly randomly chosen individual updates its opinion by partially mimicking the opinion of a uniformly randomly chosen neighbor. We numerically find that the characteristic time to the convergence increases as the system size increases according to a particular functional form in the case of lattice networks. In contrast, unless the individuals perfectly copy the opinion of their neighbors in each opinion updating, the convergence time is approximately independent of the system size in the case of regular random graphs, uncorrelated scale-free networks, and complete graphs. We also provide a mean-field analysis of the model to understand the case of the complete graph. 
\end{abstract}

\maketitle


\section{Introduction}\label{sec:intro}

In the last few decades, social dynamics has been extensively studied in various research fields including statistical physics and complex systems~\cite{Castellano2009Statistical, Sen2014Sociophysics}. A main drive underlying such studies is that many social phenomena may be understood in terms of complex macroscopic patterns that emerge from the interaction between microscopic constituents. In the present study, we focus on opinion dynamics models, which have contributed to  understanding how collective opinion evolves in a society of individuals who learn from their neighbors as well as from other sources of information such as media~\cite{Castellano2009Statistical, Acemoglu2011Opinion, Sen2014Sociophysics, Sirbu2017Opinion, Proskurnikov2017Tutorial, Proskurnikov2018Tutorial, Baronchelli2018Emergence, Anderson2019Recent, Noorazar2020Recent}. In these models, the structure of interaction among individuals has mainly been modeled by graphs, equivalently networks, in which nodes and edges represent individuals and their pairwise interaction, respectively~\cite{Borgatti2009Network, Barabasi2016Network, Newman2018Networks, Menczer2020First}.

Most models of opinion dynamics have assumed that the opinion of each individual is either a discrete or continuous variable. The prototypical model with discrete opinions is the voter model in which each individual takes one of the two opinions at any given time~\cite{Clifford1973Model, Holley1975Ergodic, Liggett1999Stochastic, Redner2019Realityinspired}. Multistate voter models are variants of the voter model in which individuals take one of more than two opinions~\cite{Howard1998Persistence, Starnini2012Ordering, Pickering2016Solution, Vazquez2019Multistate}. Other variants include multistate opinion dynamics coevolving with the network structure~\cite{Holme2006Nonequilibrium, Kimura2008Coevolutionary, Herrera2011General, Bohme2012Fragmentation} and multistate majority-vote models~\cite{Brunstein1999Universal, Tome2002Cumulants, Chen2005Consensus, Melo2010Phase, Li2016Discontinuous, Chen2018Phase}. Opinion dynamics models with continuous opinions may be defined with or without bounded confidence~\cite{Deffuant2000Mixing, Hegselmann2002Opinion, Lorenz2007Continuous, Castellano2009Statistical}. Models with bounded confidence, such as the Deffuant-Weisbuch model~\cite{Deffuant2000Mixing} and the Hegselmann-Krause model~\cite{Hegselmann2002Opinion}, assume that individuals interact with each other only when their opinions are close enough. Continuous-opinion models without the bounded confidence, which we consider in the present study, include the model by Abelson~\cite{Abelson1964Mathematical}, the DeGroot model~\cite{Degroot1974Reaching}, the Friedkin-Johnsen model~\cite{Friedkin1990Social}, and their variants. A majority of opinion dynamics models explored in control theory research community are continuous-opinion models as well~\cite{Boyd2006Randomized, Olfati-Saber2007Consensus, Fagnani2008Randomized, Nowzari2019Eventtriggered}.

A main concern in opinion dynamics models is the emergence of opinion clusters through agreement, compromise, or imitation processes, starting from initially random or diverse opinions~\cite{Castellano2009Statistical, Acemoglu2011Opinion, Sen2014Sociophysics, Sirbu2017Opinion, Proskurnikov2017Tutorial, Proskurnikov2018Tutorial, Baronchelli2018Emergence, Anderson2019Recent, Noorazar2020Recent}. When there is a unique opinion cluster in the stationarity, it is called the consensus. In the consensus, all individuals share the same opinion. The time needed to reach the consensus, i.e., the consensus time, is known to depend on the system size, i.e., the number of individuals. Relationships between the consensus time and the system size have been studied for the voter model on various networks~\cite{Cox1989Coalescing, ben-Abraham1990Saturation, Vilone2004Solution, Sood2005Voter, Suchecki2005Voter, Suchecki2005Conservation, Castellano2005Comparison, Castellano2005Effect, Sood2008Voter, Vazquez2008Analytical, Iwamasa2014Networks, Masuda2014Voter}. For continuous-opinion models, perfect consensus would require an infinite amount of time, but one can define the convergence time to consensus in multiple reasonable manners. In particular, when a continuous-opinion dynamics is driven by an operator matrix, such as the Laplacian matrix, the convergence time has often been investigated in terms of the relevant eigenvalue of the matrix, such as the spectral gap of the Laplacian matrix~\cite{Boyd2006Randomized, Almendral2007Dynamical, Aysal2009Broadcast, Acemoglu2011Opinion, Masuda2013Temporal, Chen2021Convergence}. However, these lines of research mainly focus on the dependence of the convergence time on the structure of the network of the same size rather than on that on the system size. The dependence of the consensus time on the system size in continuous-opinion models has been studied but mostly when the consensus is achievable in finite time, i.e., when interacting individuals end up having the same opinion, e.g., the average of their opinions as in some gossip models~\cite{Shi2016FiniteTime, Ayiad2017Agreement, Kouachi2020Convergence}. Therefore, the dependence of the convergence time on the system size in continuous-opinion models when the perfect consensus requires infinite time has not yet been thoroughly explored.

In the present study, we numerically investigate finite-size effects on the time towards consensus in a simple model of continuous-opinion dynamics without a bounded confidence, namely, the asymmetric-gossip model that was proposed in previous studies~\cite{Fagnani2008Randomized, Acemoglu2011Opinion}. For this purpose, we consider networks with different numbers of nodes and structure. In the asymmetric-gossip model, a uniformly randomly chosen individual updates its opinion by partially mimicking the opinion of a uniformly randomly chosen neighbor. Thus, this model can be considered as a continuous-opinion version of the voter model.

\section{Model}\label{sec:model}

We study a continuous-opinion dynamics model proposed in Refs.~\cite{Fagnani2008Randomized, Acemoglu2011Opinion}. In this model, which we refer to as the asymmetric-gossip model following Ref.~\cite{Fagnani2008Randomized}, the opinion of each individual is represented by a real value. For a system of $N$ individuals, we denote the opinion of the $i$th individual ($i=1,\ldots,N$) at time $t$ by $x_i(t) \in [0,1]$. The individuals interact on a connected network of $N$ nodes, and an individual is located at each node. There are $N$ attempts of opinion updating per unit time, which implies that each individual updates its opinion once per unit time on average. In each attempt of opinion updating, we first select an individual $i$ with the equal probability, i.e., $1/N$, and then select one of $i$'s neighbors, say $j$, uniformly at random. Then, the $i$'s opinion approaches the $j$'s opinion depending on a learning rate parameter $q$ ($0\leq q\leq 1$) as follows:
\begin{equation}
  \label{eq:update_x}
  x_i\left(t+\frac{1}{N}\right)=(1-q) x_i(t) + q x_j(t),
\end{equation}
which implies that the $i$'s new opinion is a weighted sum of the $i$'s old opinion and the $j$'s opinion. The $j$'s opinion remains the same.

If $q=0$, the opinions never change over time. If $q=1$, the model reduces to the multistate voter model~\cite{Starnini2012Ordering, Pickering2016Solution}. When $q=1$, in finite networks (i.e., $N < \infty$), the consensus in which all individuals share the same opinion is always reached in finite time. Finally, if $0<q<1$, the individuals are expected to converge to a single opinion, but in a manner different from that for the multistate voter model.

For describing the ordering dynamics of the individuals' continuous opinions when $0<q<1$, we measure the difference between the opinions of two neighboring individuals $i$ and $j$ at time $t$ defined by
\begin{equation}
  \rho_{ij}(t) \equiv |x_i(t)-x_j(t)|.
\end{equation}
We define the network-level difference in opinion by
\begin{equation}
  \label{eq:ordering}
  \rho(t) \equiv \frac{1}{|E|}\sum_{(i, j)\in E}\rho_{ij}(t),
\end{equation}
where $E$ is the set of edges of the network, and $|E|$ is the number of edges. In numerical simulations, we measure the convergence time, denoted by $T$, which we define as the time at which $\rho(t)$ becomes smaller than $10^{-10}$ for the first time.

In each simulation, we draw the initial opinion of each individual $i$ uniformly randomly from the interval $[0,1]$.

\begin{figure}[!t]
  \includegraphics[width=\columnwidth]{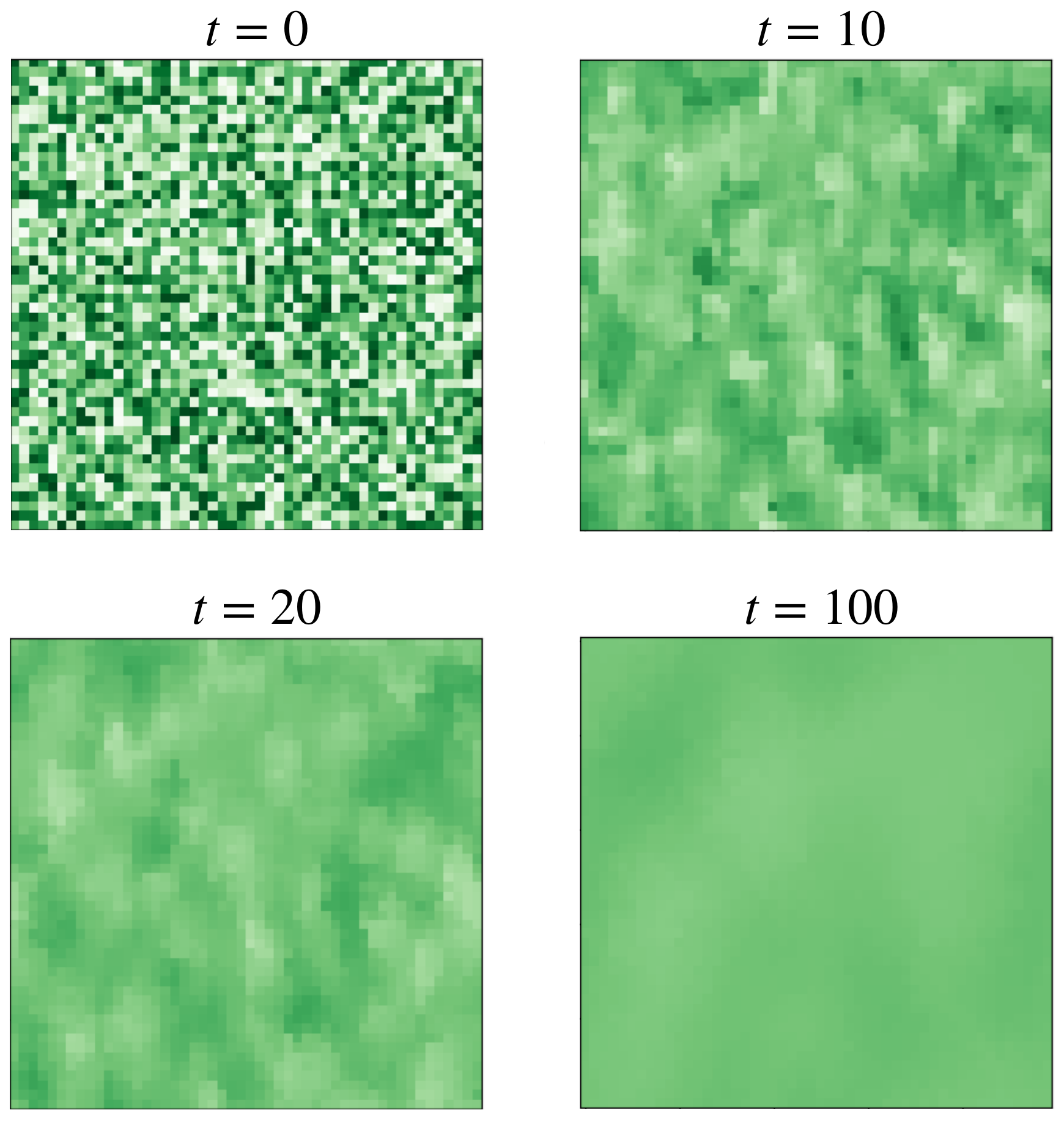}
  \caption{Snapshots at different times of the dynamics of the asymmetric-gossip model on the two-dimensional lattice. We assume $N=50\times 50$ nodes and the periodic boundary conditions. We set $q=1/2$. The opinion of a larger value is colored in a darker green. We implemented the visualization using the PyCX project~\cite{Sayama2013PyCX}.}
  \label{fig:snapshot}
\end{figure}

\begin{figure*}[!t]
  \includegraphics[width=0.8\textwidth]{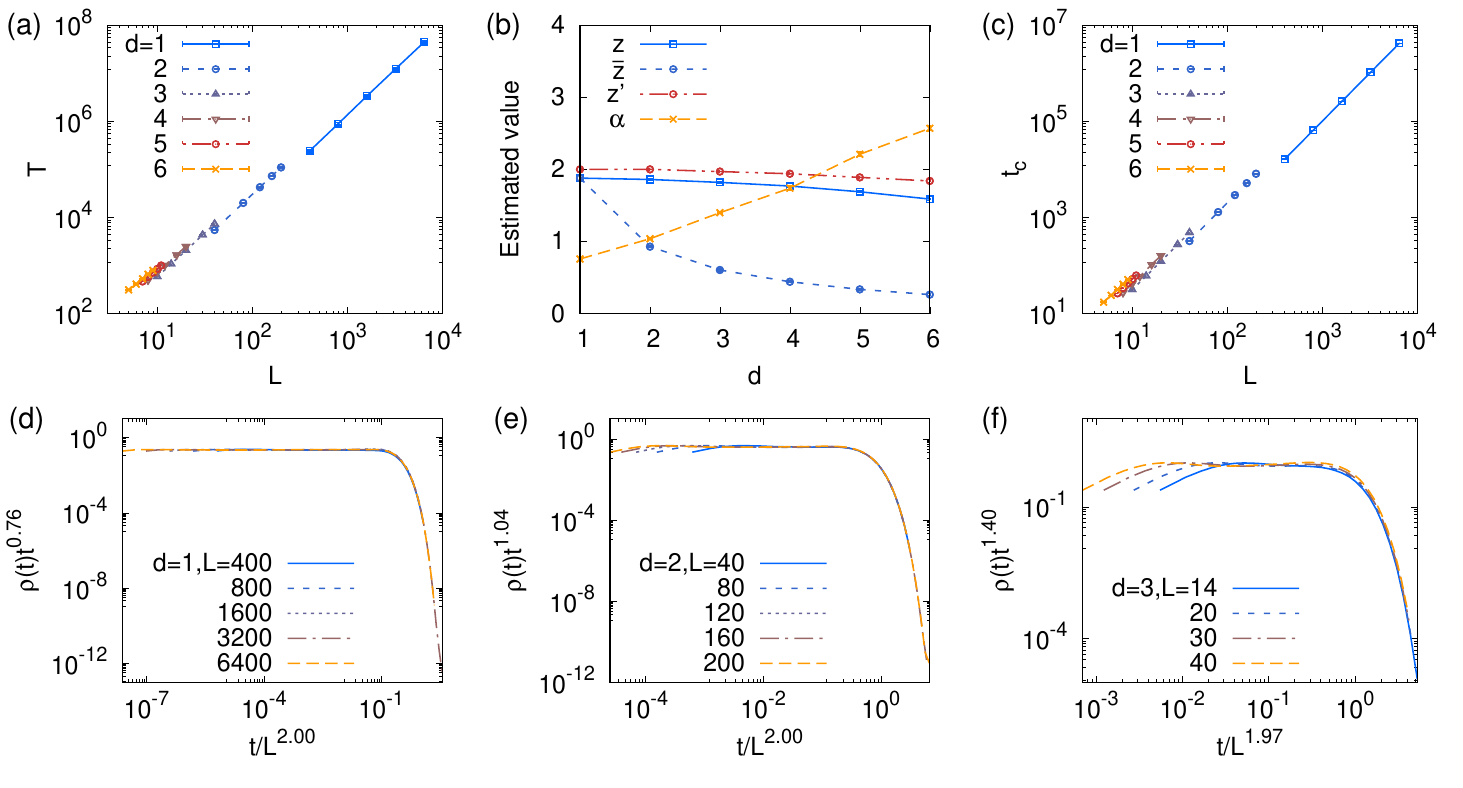}
  \caption{Simulation results of the asymmetric-gossip model on $d$-dimensional lattices with $N=L^d$ nodes under periodic boundary conditions. We set $q=1/2$. Each point or curve in panels (a), (d), (e), and (f) is an average over $100$ simulations. We show the scaling behavior of $T$ versus $L$ for $d=1,\ldots,6$ (a), the estimated values of $z$, $z'$, and $\alpha$, together with $\bar z= z/d$, for $d=1, \ldots, 6$ (b), the scaling behavior of $t_\text{c}$  versus $L$ for $d=1,\ldots,6$ (c), and data collapse of $\rho(t)$ using the estimated values of $z'$ and $\alpha$ for $d=1$ (d), $2$ (e), and $3$ (f). In panels (a), (b), and (c), the error bars, representing the standard deviation, are smaller than the symbols.}
  \label{fig:lattice}
\end{figure*}

\section{Results}\label{sec:result}

\subsection{Numerical results}\label{subsec:numeric}

We first numerically study the asymmetric-gossip model on finite-dimensional integer lattices with periodic boundary conditions, which we simply refer to as the lattice in the following text, random regular graphs (RRGs), and uncorrelated scale-free networks (SFNs).

\subsubsection{Lattices}

Let us consider the asymmetric-gossip model on the $d$-dimensional lattices with linear size $L$, combined with a periodic boundary condition. The network contains $N=L^d$ nodes. We show a typical time course of the model on the two-dimensional lattice with $q=1/2$ in Fig.~\ref{fig:snapshot}. We show the relationship between the convergence time, $T$, and $L$ for $q=1/2$ and $d=1, \ldots, 6$ in Fig.~\ref{fig:lattice}(a). The figure suggests the relationship $T\propto L^z$, where $\propto$ denotes ``proportional to''. Then, for each dimension $d$, we estimate the value of $z$ by a linear fit between $\ln T$ and $\ln L$, as shown by the solid lines in Fig.~\ref{fig:lattice}(a). See Appendix~\ref{app:expo} for details of the fitting procedure.

We show the estimated values of $z$ by the solid line in Fig.~\ref{fig:lattice}(b). The figure suggests up to $d=6$ that $z$ slightly decreases from $2$ as $d$ increases. The deviation of $z$ from $2$ for higher $d$ might be due to finite-size effects; for example, the largest linear size $L$ in our numerical simulations is only $9$ when $d=6$. The $z$ value being close to $2$ can be related to the dynamic exponent for the normal diffusion (Appendix~\ref{app:diffusion}). To later compare the present results with those for other networks, we define another exponent $\bar z$ by
\begin{equation}
    T\propto N^{\bar z}.
\end{equation}
Because $N=L^d$, one obtains 
\begin{equation}
    \bar z=\frac{z}{d}.
    \label{eq:zbar}
\end{equation}
Exponent $\bar z$ decreases as $d$ increases [see Fig.~\ref{fig:lattice}(b)]. 

To further characterize dynamics towards convergence, we numerically examine $\rho(t)$ for the same variety of the values of $d$ and $L$. In all cases $\rho(t)$ algebraically decays as a function of time before it starts to decay exponentially. Therefore, we assume that
\begin{equation}
  \rho(t)\propto t^{-\alpha}e^{-t/t_\text{c}},\ t_\text{c}\propto L^{z'},
\end{equation}
where $\alpha$ is the decay exponent, and $z'$ is the dynamic exponent relating the characteristic time $t_\text{c}$ and the linear size $L$. Similar to the estimation of $z$, for each dimension $d$, we estimate the value of $z'$ by a linear fit between $\ln t_\text{c}$ and $\ln L$ applied to the numerical results, as shown in Fig.~\ref{fig:lattice}(c).  Then, we estimate the value of $\alpha$ from $\rho(t)$ obtained for the largest value of $L$ that we use. See Appendix~\ref{app:expo} for details of the fitting procedure.

We show the estimated values of $z'$ and $\alpha$ for $d=1,\ldots,6$ in Fig.~\ref{fig:lattice}(b). The figure suggests up to $d=6$ that $\alpha$ increases as $d$ increases and that the behavior of $z'$ is quantitatively similar to that of $z$. For each dimension, with the estimated values of $z'$ and $\alpha$, we find that the curves of $\rho(t)$ for different values of $L$ collapse onto a single curve when we plot $\rho(t)t^{\alpha}$ versus $t/L^{z'}$, as shown in Figs.~\ref{fig:lattice}(d)--(f).

We contrast these findings with the results for the voter model with two discrete opinions~\cite{Krapivsky2010Kinetic}. For the voter model, the density of edges connecting nodes with the opposite opinions, denoted by $\rho_\text{a}(t)$, decays as $\rho_\text{a}(t)\propto t^{-1/2}$ for $d=1$, decays as $\rho_\text{a}(t)\propto(\ln t)^{-1}$ for $d=2$, and converges to a positive constant for $d>2$. The consensus time, denoted by $\overline{T}$, for the voter model in finite networks scales as $\overline{T}\propto N^2$ for $d=1$, $\overline{T}\propto N\ln N$ for $d=2$, and $\overline{T}\propto N$ for $d>2$, implying that $\bar z=2$ for $d=1$ and $\bar z=1$ for $d\ge 2$. The mean-field theory for the original and multistate voter models in finite networks also yields $\bar z=1$~\cite{Vazquez2008Analytical, Starnini2012Ordering}. The numerically estimated values of $\bar z$ for the asymmetric-gossip model and the result for the voter model are consistent with each other for $d\le 2$, but they are different for $d>2$. We also remark that the Ising model and a variant of the voter model called the confident voter model on two-dimensional lattices often show the metastable structure (e.g., stripes) that induces two different timescales for relaxation; see Refs.~\cite{Spirin2001Fate, Spirin2001Freezing} for the Ising model and Ref.~\cite{Volovik2012Dynamics} for the confident voter model. The present asymmetric-gossip model does not show such metastable states.

\begin{figure}[!t]
  \includegraphics[width=\columnwidth]{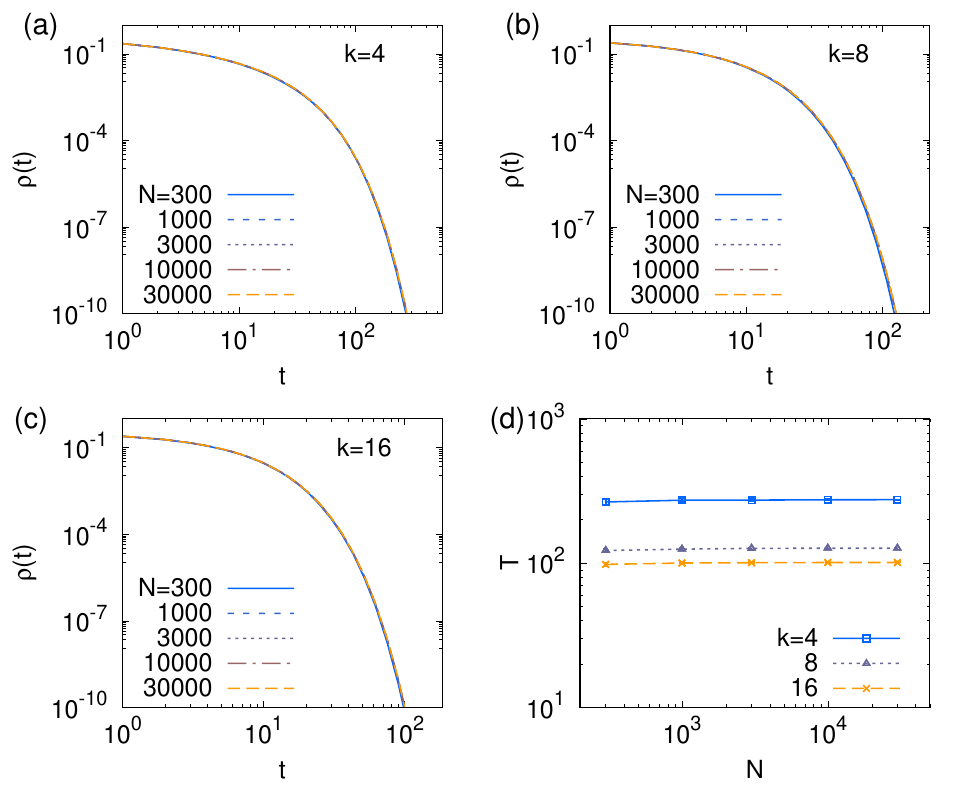}
  \caption{Simulation results of the asymmetric-gossip model on random regular graphs (RRGs) with $N$ nodes and degree $k$ for each node. We set $q=1/2$. Each point on the curve is an average over $100$ simulations on a single realization of the RRG. We show $\rho(t)$ for various network sizes with $k=4$ (a), $8$ (b), and $16$ (c). The convergence time is plotted in panel (d), where the error bars represent the standard deviation.}
  \label{fig:RRG}
\end{figure}

\subsubsection{Regular random graphs}

Next, we perform the numerical simulations of the asymmetric-gossip model on the RRGs. The numerical results in the case of $q=1/2$ and the node's degree $k=4$, $8$, and $16$ are shown in Figs.~\ref{fig:RRG}(a),~\ref{fig:RRG}(b), and~\ref{fig:RRG}(c), respectively. We find that $\rho(t)$ is independent of $N$ for each $k$ value. Therefore, $T$ is independent of $N$, i.e., $\bar z=0$ [see Fig.~\ref{fig:RRG}(d)]. As is the case for various complex networks, RRGs are considered to have $d\to\infty$. Therefore, the result that $\bar z=0$ is consistent with that for the lattices, for which $\bar z$ decreases as the dimension of the lattice, $d$, increases, roughly according to $\bar z \approx 2/d$ [see Fig.~\ref{fig:lattice}(b)]. Our result that $\bar z=0$ is again different from the known result for the voter model on uncorrelated networks including RRGs, i.e., $\bar z=1$~\cite{Vazquez2008Analytical}.

\begin{figure}[!t]
  \includegraphics[width=\columnwidth]{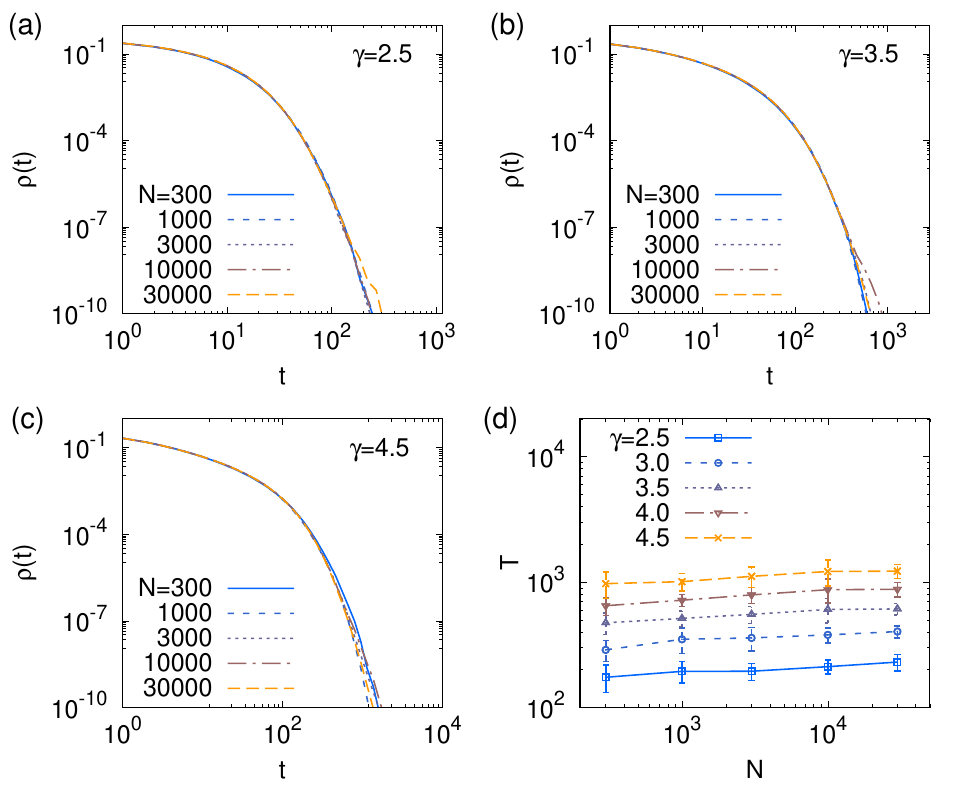}
  \caption{Simulation results of the asymmetric-gossip model on uncorrelated scale-free networks (SFNs) with degree distribution $P(k)\propto k^{-\gamma}$. We set $q=1/2$. Each point on the curve is an average over $100$ simulations. For each pair of the $N$ and $\gamma$ values, we generate $50$ SFNs to perform two simulations on each of them. We show $\rho(t)$ for various network sizes with $\gamma=2.5$ (a), $3.5$ (b), and $4.5$ (c). The convergence time is plotted in panel (d), where the error bars represent the standard deviation.}
  \label{fig:SFN}
\end{figure}

\subsubsection{Scale-free networks}

In this section, we study the asymmetric-gossip model on the uncorrelated SFNs. We assume that the node's degree obeys a power-law distribution given by $P(k)=C k^{-\gamma}$ for $k\geq k_{\rm min}=2$ with the degree exponent $\gamma$ and normalization constant $C$. 
We generate the SFNs based on the uncorrelated configuration model~\cite{Catanzaro2005Generation}. In other words, we begin with $N$ isolated nodes to each of which we assign degree $k_i$ (for $i=1, \ldots, N$) that is drawn from the distribution $P(k)$. Then, conditioned that $\sum_{i=1}^N k_i$ is an even number, we select a pair of nodes with probabilities proportional to their remaining degrees, i.e., $k_i$ subtracted by the present degree, and connect them by an edge if both of their current degrees are smaller than the assigned degrees and there is no edge between them. We repeat this wiring procedure until all nodes have the actual degree $k_i$. In practice, we terminate the wiring procedure when nodes with positive remaining degrees are already connected to each other or when there is only one node with the positive remaining degree. The generated networks are connected.

The numerical results on SFNs are shown in Fig.~\ref{fig:SFN} for $q=1/2$ and for several values of $\gamma$ and $N$. We find that both $\rho(t)$ and $T$ are only slightly affected by $N$, which implies that $\bar z\approx 0$, for a range of $\gamma$.

For the voter model on SFNs, the consensus time scales as $\overline{T}\propto N$ for $\gamma>3$, $\overline{T}\propto N/\ln N$ for $\gamma=3$, $\overline{T}\propto N^{(2\gamma-4)/(\gamma-1)}$ for $2<\gamma<3$, $\overline{T}\propto (\ln N)^2$ for $\gamma=2$, and $\overline{T}\propto \mathcal{O}(1)$ for $\gamma<2$~\cite{Sood2005Voter}. Therefore, $\bar z$ is dependent on $\gamma$ and positive for the voter model if $\gamma > 2$. In contrast, we obtain $\bar z \approx 0$ for any $\gamma \ge 2.5$ in the asymmetric-gossip model.

\begin{figure}[!t]
  \includegraphics[width=\columnwidth]{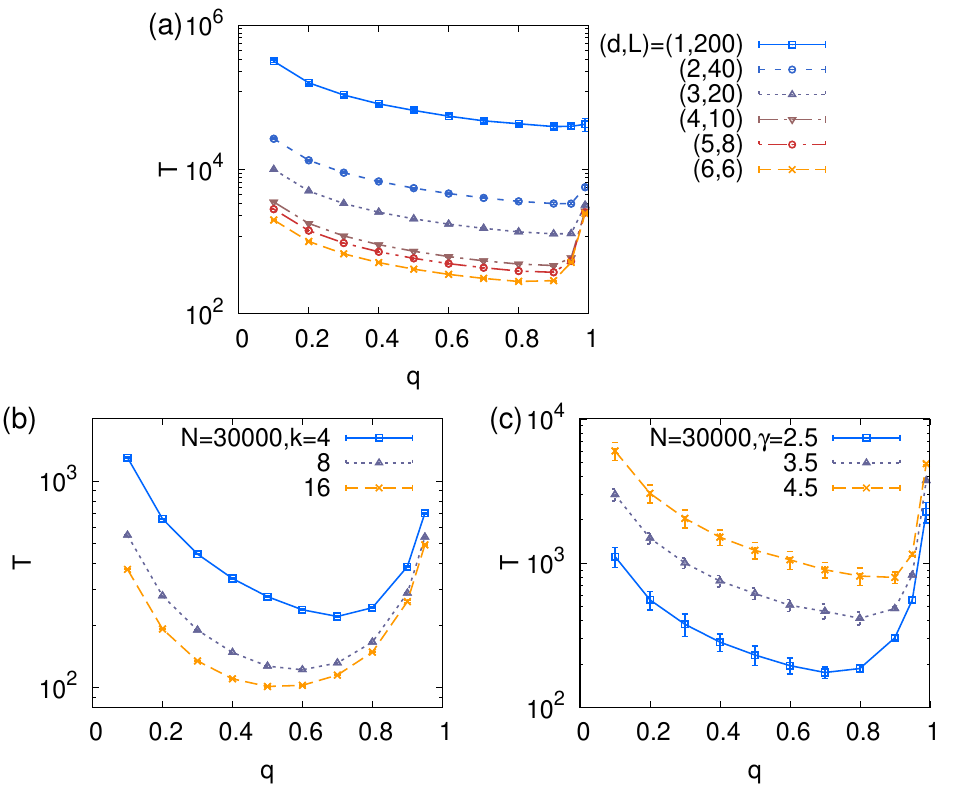}
  \caption{Convergence time, $T$, of the asymmetric-gossip model on the different networks as a function of $q$. (a) $d$-dimensional lattices with linear size $L$. (b) Regular random graphs with degree $k$. (c) Uncorrelated scale-free networks with degree exponent $\gamma$. These networks are the same as those used in Figs.~\ref{fig:lattice}--\ref{fig:SFN}. Each symbol is an average over $100$ simulations. The error bars represent the standard deviation.}
  \label{fig:q}
\end{figure}

\subsubsection{Effect of the learning rate parameter $q$}

We now study the effect of the learning rate parameter $q$, introduced in Eq.~\eqref{eq:update_x}. We show the convergence time, $T$, as a function of $q$ for the lattices, RRGs, and SFNs of different sizes in Fig.~\ref{fig:q}. In all cases, we find the optimal value of $q$ between $0$ and $1$ in terms of the convergence speed, making $T$ the smallest. These results imply that the faster learning does not always speed up the convergence.

\begin{figure*}[!t]
  \includegraphics[width=0.8\textwidth]{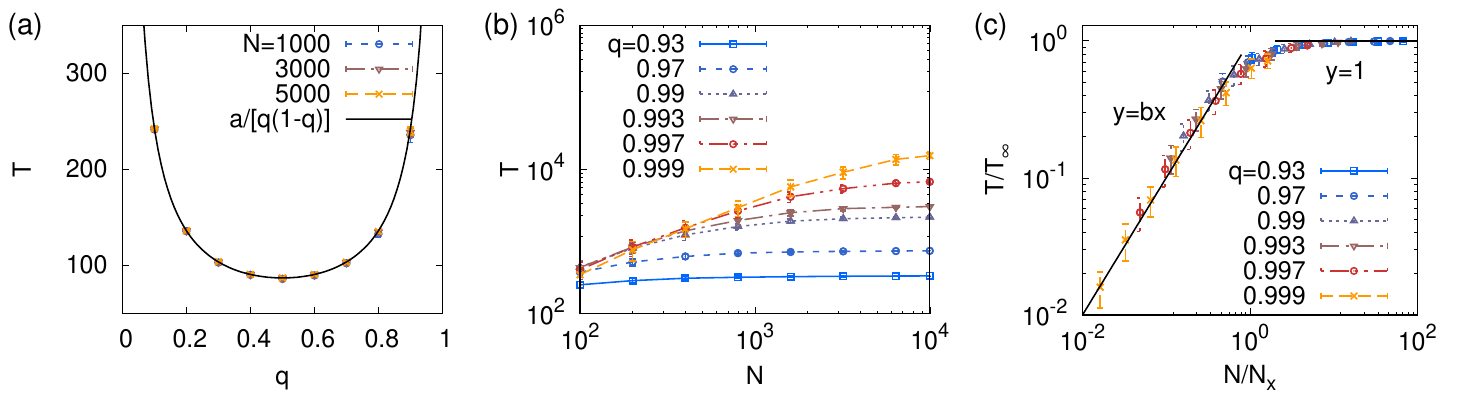}
  \caption{Convergence time for the asymmetric-gossip model on complete graphs. Each symbol is an average over $100$ simulations. (a) Convergence time, $T$, as a function of $q$ for $N=10^3$, $3\times 10^3$, and $5\times 10^3$. The solid curve represents the analytical form of $T$, i.e., $a/[q(1-q)]$, where $a$ is a constant, which we set to $21.8$; see Eq.~\eqref{eq:tau_c_MF}. (b) Convergence time as a function of $N$ for various values of $q$ close to $1$. (c) Data collapse of $T/T_\infty$ as a function of $N/N_\times$ based on the results shown in panel (b). We set $T_\infty = a/[q(1-q)]$ and $N_\times=T_\infty/b$, where $b = 3.55$. In all panels, the error bars represent the standard deviation. The solid lines in (c) represent $y=1$ and $y=bx$.}
  \label{fig:MF}
\end{figure*}

\subsection{Mean-field analysis}\label{subsec:anal}

To understand our numerical results, in this section we analyze the mean-field case where every node equally interacts with every other node. The state of the system at time $t$ is specified by the distribution of opinions, $\{x_i(t)\}$, which we denote by $P_t(x)$. By using Eq.~\eqref{eq:update_x} and approximating the original discrete time by continuous time, which corresponds to the limit $N\to\infty$, we obtain the following master equation to describe the dynamics of $P_t(x)$:
\begin{eqnarray}
  \frac{\partial P_t(x)}{\partial t} &=& \int_0^1 \text{d}x_i \int_0^1 \text{d}x_j P_t(x_i)P_t(x_j) \nonumber\\
  &\times & \left[\delta(x-(1-q)x_i-qx_j) - \delta(x-x_i)\right],
  \label{eq:master_origin}
\end{eqnarray}
where $\delta$ denotes Dirac delta. In Eq.~\eqref{eq:master_origin}, $x_i$ and $x_j$ are the opinions of a uniformly randomly chosen node $i$, whose opinion changes at time $t$, and another uniformly randomly chosen node $j$, respectively. Equation~\eqref{eq:master_origin} is similar to the equation that has originally been proposed for representing inelastic collisions of particles~\cite{Ben-Naim2003Bifurcations, Ben-Naim2000Multiscaling}. By taking the Laplace transform of Eq.~\eqref{eq:master_origin} with respect to $x$ over the range $[0,1]$, one obtains
\begin{equation}
  \label{eq:master_s}
  \frac{\partial \tilde P_t(s)}{\partial t} =-\tilde P_t(s)+\tilde P_t((1-q)s) \tilde P_t(qs),
\end{equation}
where 
\begin{equation}
  \label{eq:Laplace}
  \tilde P_t(s)\equiv \int_0^1 \text{d}x e^{-xs} P_t(x).
\end{equation}
We expand the exponential term in Eq.~\eqref{eq:Laplace} to obtain
\begin{equation}
  \label{eq:expand_exp}
  \tilde P_t(s) = \sum_{n=0}^\infty \frac{(-s)^n}{n!} c_n(t),
\end{equation}  
where 
\begin{equation}
    c_n(t) \equiv \int_0^1 \text{d}x x^n P_t(x)
\end{equation}
is the $n$th moment of $x$ at time $t$. Note that $c_0(t)=1$ for any $t$. The initial uniform distribution of the opinions implies $P_0(x)=1$ for $x\in [0,1]$, which translates into $c_1(0)=1/2$. By plugging Eq.~\eqref{eq:expand_exp} into Eq.~\eqref{eq:master_s} and comparing the coefficients of $s$ and $s^2$, we obtain
\begin{eqnarray}
  \frac{d c_1(t)}{dt} &=& 0,\\
  \frac{d c_2(t)}{dt} &=& -2q(1-q)c_2(t) + \frac{q(1-q)}{2}.
  \label{eq:dc_2(t)/dt}
\end{eqnarray}
For the second equation, we have used the fact that $c_1(t)=c_1(0)=1/2$.

Instead of using $\rho(t)$ given by Eq.~\eqref{eq:ordering}, one can also describe the ordering dynamics in terms of the temporal evolution of the variance of opinions, i.e.,
\begin{eqnarray}
v(t) &\equiv& \langle x(t)^2\rangle -\langle x(t)\rangle^2\nonumber\\
&=& c_2(t)-c_1(t)^2\nonumber\\
&=& c_2(t) - \frac{1}{4}.
\label{eq:v(t)}
\end{eqnarray}
By substituting Eq.~\eqref{eq:v(t)} into Eq.~\eqref{eq:dc_2(t)/dt}, we obtain
\begin{equation}
  \frac{d v(t)}{dt} = -2q(1-q)v(t).
\end{equation}
Therefore, for $0<q<1$ we obtain
\begin{equation}
  \label{eq:MF_var}
  v(t)=v(0)e^{-t/\tau_\text{c}},
\end{equation}
where
\begin{equation}
\tau_\text{c}\equiv \frac{1}{2q(1-q)}.
\label{eq:tau_c_MF}
\end{equation}
Therefore, the opinion distribution, $P_t(x)$, converges to $\delta(x-1/2)$, i.e., the consensus at $x=1/2$, exponentially fast. Equation~\eqref{eq:tau_c_MF} implies that the convergence time, $T$, is also proportional to $\left[q(1-q)\right]^{-1}$, which we numerically confirm for complete graphs [see Fig.~\ref{fig:MF}(a)]. Note that the convergence time, $T$, for the complete graph is independent of the number of nodes, $N$, for $q=0.1,\ldots,0.9$, which we used in Fig.~\ref{fig:MF}(a).

\subsection{Crossover behavior}\label{subsec:cross}

We have found for RRGs and SFNs that $\bar z \approx 0$ for $0<q<1$. However, we expect to find $\bar z=1$, i.e., $T\propto N$ for $q=1$ because when $q=1$, the asymmetric-gossip model reduces to the multistate voter model. Therefore, we expect to find a crossover between the $q<1$ regime and the $q=1$ case in terms of the dependence of $T$ on $N$. To study this crossover behavior, we carry out numerical simulations on the complete graph with several values of $q$ close to one. 

We show the relationship between $T$ and $N$ in Fig.~\ref{fig:MF}(b). The figure suggests for each $q$ that $T\propto N$ for $N\ll N_\times$ and that $T$ is constant for $N\gg N_\times$, where $N_\times$ denotes a crossover system size. Therefore, we assume that
\begin{equation}
    T=T_\infty f\left(\frac{N}{N_\times}\right),
    \label{eq:scaling_form_T}    
\end{equation}
where
\begin{equation}
    f(r)=\begin{cases}
    r & \textrm{if}\ r\ll 1,\\
    1 & \textrm{if}\ r\gg 1.
    \end{cases}
    \label{eq:scaling_fn}
\end{equation}
Note that $N_\times$ may depend on $q$. If $N\gg N_\times$, one obtains $T=T_\infty$, which implies that $T_\infty$ is the convergence time for large $N$. In Section~\ref{subsec:anal}, we suggested $T_{\infty} \propto [q(1-q)]^{-1}$ based on the mean-field theory. Therefore, we assume that $T_{\infty} = a/[q(1-q)]$ with $a=21.8$ [see also Fig.~\ref{fig:MF}(a)]. If $N\ll N_\times$, then Eqs.~\eqref{eq:scaling_form_T} and \eqref{eq:scaling_fn} imply that $T=(T_\infty/N_\times)N\equiv bN$. We estimate $b=3.55$ by the linear fit between $T$ and $N$ using numerical results obtained for $N=100, \ldots, 1600$ with $q=0.999$. Then we write $N_\times$ as follows:
\begin{equation}
    N_\times = \frac{T_\infty}{b}=\frac{a}{bq(1-q)}.
\end{equation}
Consequently, as $q$ approaches one, $N_\times$ diverges. Using these $T_\infty$ and $N_\times$ values, which depend on $q$, we plot $T/T_\infty$ as a function of $N/N_\times$ for various values of $q$ in Fig.~\ref{fig:MF}(c). We find that all curves roughly collapse onto a single curve that satisfies Eq.~\eqref{eq:scaling_fn}.

We remark that the crossover behavior between different $N$-dependences of the consensus time $\overline{T}$ has been reported for other opinion dynamics models with discrete opinions: $\overline{T}\propto N^{1.6}$ for $N< N_\times$ and $\overline{T}\propto N^{0.2}$ for $N > N_\times$ on the one-dimensional lattice, and $\overline{T}\propto N^{0.6}$ for $N< N_\times$ and $\overline{T}\propto N^{0.1}$ for $N> N_\times$ on the two-dimensional lattice~\cite{Tessone2004Neighborhood, Toral2007Finite}.

\section{Conclusion}\label{sec:conclusion}

We have studied an asymmetric-gossip model for continuous-opinion dynamics without a bounded confidence. We have specifically focused on the effects of the system size, $N$, i.e., the number of individuals, on the convergence time, $T$. We numerically find that the scaling exponent relating $T$ and $N$ depends on the dimension of the lattices. In contrast, $T$ is approximately independent of $N$ for the regular random graphs, uncorrelated scale-free networks, and complete graphs. We have presented a mean-field analysis to support the numerical results for the complete graphs.

In Ref.~\cite{Aysal2009Broadcast}, the authors analyzed the convergence time of a model similar to the present asymmetric-gossip model. Their model is a broadcast asymmetric-gossip model. In other words, when a node $i$ updates its state according to Eq.~\eqref{eq:update_x}, all the nodes adjacent to node $j$ do so at the same time such that $j$ broadcasts its opinion to all its neighbors. They considered spatial networks to find that there is a value of $q$, where $0<q<1$, which minimizes the convergence time. This result is consistent with our numerical results for the lattices, RRGs, and SFNs [see Fig.~\ref{fig:q}] and our theoretical result that the convergence is the fastest at $q=1/2$ on the complete graph [see Eq.~\eqref{eq:tau_c_MF}]. Future questions in this direction include how our scaling results extend to broadcast gossip models, how the optimal $q$ values depend on the network structure, and how we can leverage linear algebra techniques~\cite{Fagnani2008Randomized, Aysal2009Broadcast} to further understand the present asymmetric-gossip model. 

As future work, it may also be interesting to study finite-size effects on the convergence time in more complicated, realistic models with continuous opinions, such as multidimensional Deffuant-Weisbuch and Hegselmann-Krause models~\cite{Fortunato2005Vector, Lorenz2006Continuous, Sirbu2017Opinion}.

\begin{acknowledgments}
H.-H.J. thanks Jae Dong Noh for fruitful comments on the initial draft, and acknowledges financial support by Basic Science Research Program through the National Research Foundation of Korea (NRF) grant funded by the Ministry of Education (NRF-2018R1D1A1A09081919) and by the Catholic University of Korea, Research Fund, 2020.
N.M. acknowledges support from AFOSR European Office (under Grant No. FA9550–19–1–7024), the
Nakatani Foundation, and the Sumitomo Foundation.
\end{acknowledgments}

\appendix

\begin{table*}[!ht]
\centering
\caption{Details of the simulation on lattices and the estimation of $z$, $z'$, and $\alpha$ values. For each dimension $d$, we show the linear sizes $L$, estimated value of $z$ with standard error in parentheses, fitting ranges of $t$ for $t_{\text{c}}$, estimated value of $z'$ with standard error in parentheses, fitting range of $t$ for $\alpha$, and estimated value of $\alpha$  with the standard error in parentheses.}
\begin{tabular}{c|c|c|c|c|c|c}
    \hline
    $d$ & $L$ & $z$ & Fitting range of $t$ for $t_\text{c}$ & $z'$ & Fitting range of $t$ for $\alpha$ & $\alpha$ \\
    \hline
    $1$ & $400,800,1600,3200,6400$ & 1.88(1) & $\ge 2\cdot 10^4,\ge 8\cdot 10^4,\ge 2\cdot 10^5,[10^6,3\cdot 10^7],\ge 5\cdot 10^6$ & $2.00(1)$ & $[e^{2.5},e^{15}]$  & $0.76(1)$\\
    $2$ & $40,80,120,160,200$ & $1.86(1)$ & $\ge 500,\ge 2000,\ge 5000,[10^4:14\cdot 10^4],[2\cdot 10^4:25\cdot 10^4]$ & $2.00(1)$ & $[e^{2.5},e^{9}]$  & $1.04(1)$\\
    $3$ & $10,14,20,30,40$ & $1.82(1)$ & $\ge 70,\ge 150,\ge 300,\ge 500,\ge 1000$ & $1.97(1)$ & $[e^{2.5},e^{6}]$ & $1.40(1)$ \\
    $4$ & $8,10,12,16,20$ & $1.77(1)$ & $\ge 80,\ge 100,\ge 180,\ge 250,\ge 300$ & $1.94(1)$ & $[e^{2.5},e^{5.5}]$ & $1.74(3)$ \\
    $5$ & $7,8,9,10,11$ & $1.69(1)$ & $\ge 60,\ge 70,\ge 90,\ge 100,\ge 150$ & $1.89(1)$ & $[e^{2.5},e^{5}]$  & $2.21(3)$\\
    $6$ & $5,6,7,8,9$ & $1.59(2)$ & $\ge 50,\ge 80,\ge 90,\ge 100,\ge 120$ & $1.84(2)$ & $[e^{2.5},e^{5}]$  & $2.57(4)$\\
    \hline
\end{tabular}
\label{table}
\end{table*}

\section{Estimation of $z$, $z'$, and $\alpha$ values for lattices}\label{app:expo}

On the lattice of each dimension, $d$, and each linear size, $L$, shown in Table~\ref{table}, we run $100$ simulations with different initial conditions to obtain the average convergence time $T$ as well as the average curve of $\rho(t)$. Then, we estimate the value of $z$ by the linear fit between $\ln T$ and $\ln L$, which is informed by the assumption that $T\propto L^z$. The estimated values of $z$ and their standard errors are shown in the third column of Table~\ref{table}.

From $\rho(t)$, we estimate $t_\text{c}$ by the linear fit between $\ln \rho(t)$ and $t$ for a range of $t$ showing the exponentially decaying behavior. We show the range of $t$ used for the fitting in each case in the fourth column of Table~\ref{table}. Once the values of $t_\text{c}$ for all the values of $L$ considered are ready, we estimate the value of $z'$ by the linear fit between $\ln t_\text{c}$ and $\ln L$, which is informed by the assumption that $t_\text{c}\propto L^{z'}$. The estimated values of $z'$ and their standard errors are shown in the fifth column of Table~\ref{table}.

As mentioned in the main text, for each $d$, we compute the value of $\alpha$ from the $\rho(t)$ for the largest $L$. We estimate the value of $\alpha$ by the linear fit between $\ln \rho(t)$ and $\ln t$ for a range of $t$ showing the scaling behavior, which is listed in the sixth column of Table~\ref{table}. The estimated values of $\alpha$ with their standard errors are shown in the seventh column of Table~\ref{table}.

\section{Relation of the asymmetric-gossip model to the diffusion process}\label{app:diffusion}

We can interpret the update rule in Eq.~\eqref{eq:update_x} as a normal diffusion process. To show this, we expand the left-hand side of Eq.~\eqref{eq:update_x} assuming $N\gg 1$ to obtain
\begin{equation}
    \frac{dx_i(t)}{dt} = Nq[x_j(t)-x_i(t)].
\end{equation}
On one-dimensional lattices, $j$ is either $i-1$ or $i+1$ with probability $1/2$ each. Therefore, the expectation of $x_i(t)$, denoted by $E[x_i(t)]$, evolves according to
\begin{equation}
    \frac{dE[x_i(t)]}{dt} = \frac{Nq}{2}\left\{E[x_{i-1}(t)]+E[x_{i+1}(t)]-2E[x_i(t)]\right\}.
\label{eq:relation1}
\end{equation}
One can approximate Eq.~\eqref{eq:relation1} in the continuous space by the normal diffusion equation as follows:
\begin{equation}
    \frac{d E[x(y,t)]}{dt} \approx D \frac{d^2 E[x(y,t)]}{dy^2},
\label{eq:relation2}
\end{equation}
with a diffusion constant $D=Nq/2$, where $\approx$ represents ``approximately equal to''. The dynamic exponent $z$ relating the timescale and the length scale of the normal diffusion given by Eq.~\eqref{eq:relation2} is $2$. The same derivation applies to the higher-dimensional lattices with a dimension-dependent diffusion constant $D=Nq/(2d)$, where we remind that $d$ is the dimension.

%

\end{document}